\documentclass[10pt,twocolumn]{article}
\usepackage{times}

\pdfoutput=1

\newenvironment{BoldAbstract}{%
\begin{quote} \bf}
{\end{quote}}

\usepackage{amsmath}
\usepackage{amssymb}
\usepackage{amsfonts}
\usepackage{graphicx}
\usepackage{epsfig}
\usepackage{scicite}

\newcommand{\abs}[1]{\left|#1\right|}

\newcommand{\integ}[1]{\ensuremath{\int \!\! \mathrm{d}#1 \,}}
\newcommand{\integlim}[3]{\ensuremath{\int_{#1}^{#2} \!\!\! \mathrm{d}#3 \,}}

\newcommand{\mean}[1]{\ensuremath{\left\langle #1 \right\rangle}}

\renewcommand{\a}{\ensuremath{a_L}}

\newcommand{\alphai}{\ensuremath{\alpha_{\textrm{in}}}}

\newcommand{\XM}{\ensuremath{X_m}}
\newcommand{\PM}{\ensuremath{P_m}}

\newcommand{\nbar}{\ensuremath{\bar{n}}}

\newcommand{\xzp}{\ensuremath{x_0}}

\newcommand{\nL}{\ensuremath{n_L}}
\newcommand{\nM}{\ensuremath{n_m}}

\newcommand{\ket}[1]{\ensuremath{\left| #1 \right\rangle}}

\newcommand{\comm}[2]{\ensuremath{\left[#1,#2\right]}}

\newcommand{\be}{\begin{equation}}
\newcommand{\ee}[1]{\label{#1} \end{equation}}

\def\f12{\frac{1}{2}}

\newcommand\Tspace{\rule{0pt}{2.6ex}}
\newcommand\TspaceBig{\rule{0pt}{3.0ex}}
\newcommand\Bspace{\rule[-2.0ex]{0pt}{0pt}}

\long\def\symbolfootnote[#1]#2{\begingroup%
 \def\thefootnote{\fnsymbol{footnote}}\footnotetext[#1]{#2}\endgroup}

\title{Probing Planck-scale physics with quantum optics}

\author{Igor Pikovski,$^{1, 2 }$\footnotemark $\, \,$ Michael R. Vanner,$^{1, 2}$  Markus Aspelmeyer,$^{1, 2}$ M. S. Kim,$^{3}$\footnotemark $\, \,$ \v{C}aslav Brukner$^{2,4}$ \\
                \\
\normalsize{$^{1}$ Vienna Center for Quantum Science and Technology (VCQ), Boltzmanngasse 5, A-1090 Vienna, Austria,}\\
\normalsize{$^{2}$ Faculty of Physics, University of Vienna, Boltzmanngasse 5, A-1090 Vienna, Austria,}\\
\normalsize{$^{3}$QOLS, Blackett Laboratory, Imperial College London, SW7 2BW, United Kingdom,}\\
\normalsize{$^{4}$Institute for Quantum Optics and Quantum Information (IQOQI), Austrian Academy of Sciences,}\\
\normalsize{Boltzmanngasse 3, A-1090 Vienna, Austria.}
}

\date{}

\begin{document}
\twocolumn[
  \begin{@twocolumnfalse}
    \maketitle
		\begin{BoldAbstract}
 One of the main challenges in physics today is to merge quantum theory and
 the theory of general relativity into a unified framework. Researches are developing various approaches towards such a theory of quantum gravity, but a major hindrance
 is the lack of experimental evidence of quantum gravitational effects.
 Yet, the quantization of space-time itself can have experimental implications:
 the existence of a minimal length scale is widely expected to result in a modification
 of the Heisenberg uncertainty relation. Here we introduce a scheme to
 experimentally test this conjecture by probing directly the canonical commutation
 relation of the center-of-mass mode of a mechanical oscillator
 with a mass close to the Planck mass. Our protocol utilizes quantum optical
 control and readout of the mechanical system to probe possible deviations
 from the quantum commutation relation even at the Planck scale. We show
 that the scheme is within reach of current technology. It thus opens a feasible
 route for table-top experiments to explore possible quantum gravitational phenomena.
 \end{BoldAbstract}
  \end{@twocolumnfalse}
  ]

\symbolfootnote[1]{Electronic address: igor.pikovski@univie.ac.at}
\symbolfootnote[2]{Electronic address: m.kim@imperial.ac.uk}


It is currently an open question whether our underlying concepts of space-time are fully compatible with those of quantum mechanics.
The ongoing search for a quantum theory of gravity is therefore one of the main challenges in modern physics.
A major difficulty in the development of such theories is the lack of experimentally accessible phenomena that could shed light on the possible route for quantum gravity. Such phenomena are expected to become relevant near the Planck scale, i.e. at energies on the order of the Planck energy $E_P = 1.2 \times 10^{19}$~GeV or at length scales near the Planck length  $L_P = 1.6 \times 10^{-35}$~m, where space-time itself is assumed to be quantized.
However, such a minimal length scale is not a feature of quantum theory. The Heisenberg uncertainty relation, one of the cornerstones of quantum mechanics \cite{ref:Heisenberg1927}, states that the position $x$ and the momentum $p$ of an object cannot be simultaneously known to arbitrary precision.
Specifically, the indeterminacies of a joint measurement of these canonical observables are always bound by  $\Delta x \Delta p \geq \hbar /2$. Yet, the uncertainty principle still allows for an arbitrarily precise measurement of only one of the two observables, say position, on the cost of our knowledge about the other (momentum). In stark contrast, in many proposals for quantum gravity the Planck length constitutes a fundamental bound below which position cannot be defined.
It has therefore been suggested that the uncertainty relation should be modified in order to take into account such quantum gravitational effects \cite{ref:garay1994}. In fact, the concept of a generalized uncertainty principle is found in many approaches to quantum gravity, for example in string theory \cite{ref:strings1, ref:strings2, ref:strings3}, in the theory of doubly special relativity \cite{ref:Amelino2002, ref:Magueijo2003}, within the principle of relative locality \cite{ref:Amelino2011} and in studies of black holes \cite{ref:blackhole, ref:Scardigli1999, ref:Jizba2010}. A generalized uncertainty relation also follows from a deformation of the underlying canonical commutator $\comm{x}{p} \equiv xp - px$ \cite{ref:Maggiore,  ref:Kempf1995, ref:Das2008, ref:Ali2009, ref:Ali2011}, since they are related via $\Delta x \Delta p \geq \f12 | \langle \comm{x}{p} \rangle |$.

Preparing and probing quantum states at the Planck scale is beyond today's experimental possibilities. Current approaches to test quantum gravitational effects mainly focus on high-energy scattering experiments, which operate still $15$ orders of magnitude away from the Planck energy $E_P$, or on astronomical observations \cite{ref:astro1, ref:astro2}, which have not found any evidence of quantum gravitational effects as of yet \cite{ref:astro3, ref:astro4}. Another route would be to perform high-sensitivity measurements of the uncertainty relation, since any deviations from standard quantum mechanics are, at least in principle, experimentally testable \cite{ref:Das2008, ref:Ali2009, ref:Ali2011}. However, with the best position measurements being of order  $\Delta x /L_p \sim 10^{17}$ \cite{ref:GEO, ref:LIGO}, current sensitivities are still insufficient and quantum gravitational corrections remain unexplored.

Here we propose a scheme that circumvents these limitations. Our scheme allows to test quantum gravitational modifications of the canonical commutator in a novel parameter regime, thereby reaching a hitherto unprecedented sensitivity in measuring Planck-scale deformations. The main idea is to use a quantum optical ancillary system that provides a direct measurement of the canonical commutator of the center of mass of a massive object. In this way Planck-scale accuracy of position measurements is not required. Specifically, the commutator of a very massive quantum oscillator is probed by a sequence of interactions with a strong optical field in an opto-mechanical setting, which utilizes radiation pressure inside an optical cavity \cite{ref:review1, ref:review3}. The sequence of optomechanical interactions is used to map the commutator of the mechanical resonator onto the optical pulse. The optical field experiences a measurable change that depends on the commutator of the mechanical system and that is non-linearly enhanced by the optical intensity. Observing possible commutator deformations thus reduces to a measurement of the mean of the optical field, which can be performed with very high accuracy by optical interferometric techniques. We show that already with state-of-the art technology tests of Planck-scale deformations of the commutator are within experimental reach.

\section{Modified commutation relations}
\begin{figure}[t]
\includegraphics[width=1.0\columnwidth]{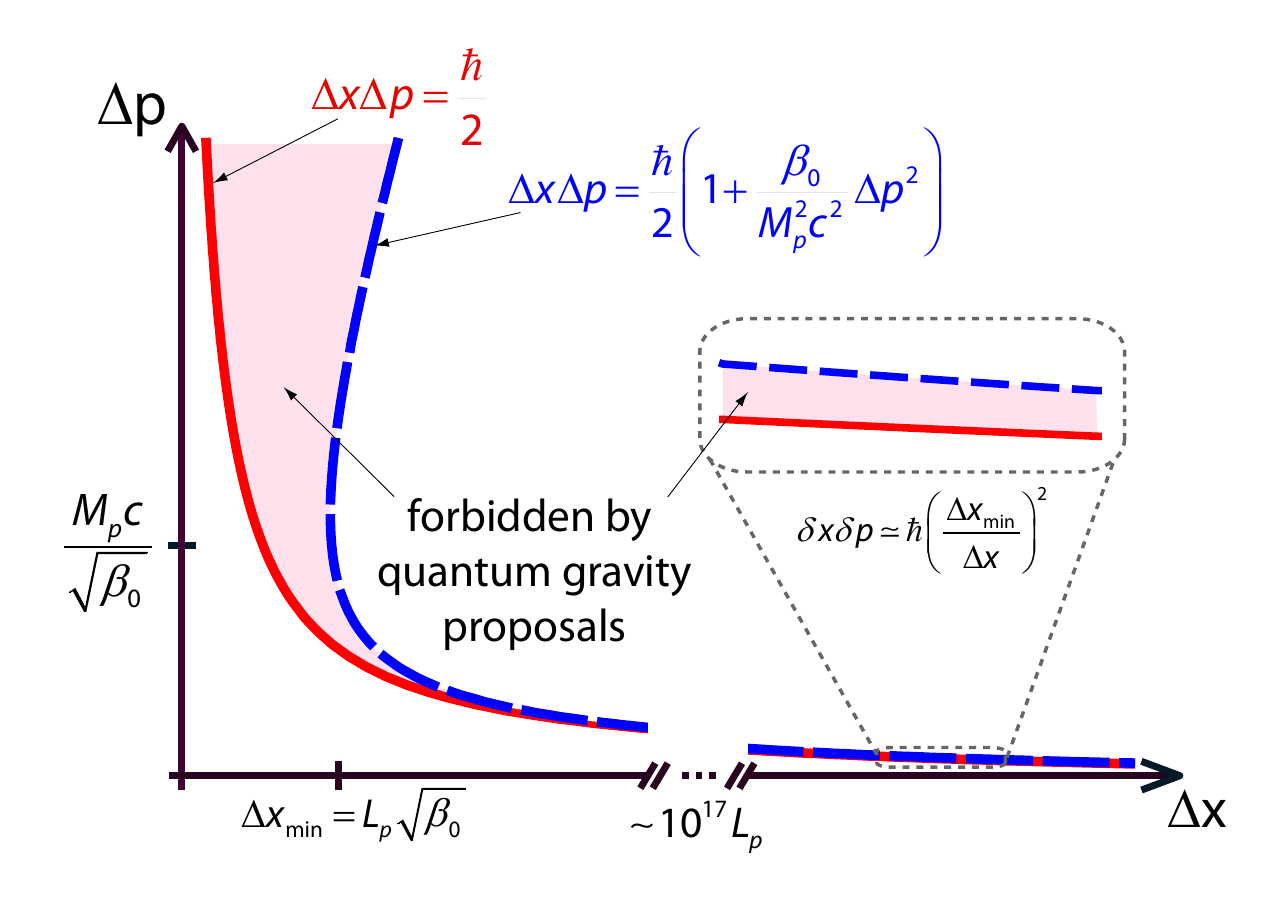}
\caption{\small The minimum Heisenberg uncertainty (red curve) is plotted together with a modified uncertainty relation (dashed blue curve) with modification strength $\beta_0$. $M_P $ and $L_P$ are the Planck mass and Planck length, respectively. The shaded region represents states that are allowed in regular quantum mechanics but are forbidden in theories of quantum gravity that modify the uncertainty relation. The inset shows the two curves far from the Planck scale at typical experimental position uncertainties $\Delta x \gg \Delta x_{min}$. An experimental precision of $\delta x$ $\delta p$ is required to distinguish the two curves, which is beyond current experimental possibilities. However, this can be overcome by our scheme that allows to probe the underlying commutation relation in massive mechanical oscillators and its quantum gravitational modifications.
} \label{fig:uncertainties}
\end{figure}

A common modification of the Heisenberg uncertainty relation that appears in a vast range of approaches to quantum gravity \cite{ref:garay1994, ref:strings1, ref:strings2, ref:Milburn2006, ref:Kempf2009} is $\Delta x \Delta p \geq \hbar \left(1+  \beta_0 \left(\Delta p / (M_{P} c)\right)^2 \right)/2$.
Here, $\beta_0$ is a numerical parameter that quantifies the modification strength, $c$ is the speed of light and $M_{P}\simeq22~\mu$g is the Planck-mass. The minimal measurable length scale appears as a natural consequence with $\Delta x_{min}= L_P \sqrt{\beta_0}$ (see Figure 1). Such a modification alters the allowed state-space and can be seen as a manifestation of a deformed canonical commutator, for example of the form \cite{ref:Kempf1995}
\be
\comm{x}{p}_{\beta_0}=i \hbar \left(1+   \beta_0 \left(\frac{p}{ M_{P} \, c} \right)^2 \right) \, .
\ee{eq:CommMod}
Up to date, no effect of a modified canonical commutator has been observed in experiments. The best currently available measurement precision (see Table \ref{table:expparamters}) allows to put an upper bound on the magnitude of the deformation of $\beta_0 < 10^{33}$  \cite{ref:Das2008}. For theories that modify the commutator this rules out the existence of an intermediate fundamental length scale on the order of $x \sim 10^{-19}$~m. Note that the Planck scale modifications correspond to $\beta_0 \sim 1$ and are therefore untested. Additionally, the above modification of the commutator is not unique and experiments can, in principle, distinguish between the various theories. In particular, a generalized version of the commutator deformation is \cite{ref:Maggiore}
\be
\comm{x}{p}_{\mu_0}=i \hbar \sqrt{1+  2 \mu_0 \frac{(p/c)^2 + m^2}{ M_{P}^2}} \, .
\ee{eq:CommMag}
Here $m$ is the mass of the particle and $\mu_0$ is again a free numerical parameter. For small masses $m \ll p /c \lesssim M_{Pl} $, and for $ \mu_0 = \beta_0$, the above modified commutator reduces to Eq. \ref{eq:CommMod}. However, an important difference is that the commutation relation in Eq. \ref{eq:CommMag} depends directly on the rest-mass of the particle. In the limit $p/c \ll m \lesssim M_{P}$ the commutator reduces to $\comm{x}{p}_{\mu_0} \approx i \hbar \left( 1+ \mu_0 m^2 / M_{P}^2\right)$, which can be seen as a mass-dependent rescaling of $\hbar$. It is worth noting that a modified, mass-dependent Planck constant $\hbar=\hbar(m)$ also appears in other theories, some of which predict that the value of Planck's constant can decrease with increasing mass ($\hbar \rightarrow 0$ for $m \gg M_{P}$), in contrast to the prediction above. Such a reduction would also account for a transition to classicality in massive systems or at energies close to the Planck energy \cite{ref:Magueijo2003, ref:Jizba2010}.

Among the various proposals for different commutator deformations, we choose as a last example the recently proposed commutator \cite{ref:Ali2009} which also accounts for a maximum momentum that is present in several approaches to quantum gravity \cite{ref:Amelino2002, ref:Magueijo2003}
\be
\comm{x}{p}_{\gamma_0}=i \hbar \left(1 -  \gamma_0 \frac{p}{ M_{P} \, c} +  \gamma_0^2 \left(\frac{p}{ M_{P} \, c} \right)^2 \right) \, .
\ee{eq:CommADV}
Here $\gamma_0$ is again a free numerical parameter that characterizes the strength of the modification. Experimental bounds on $\gamma_0$ are more stringent than in the case of Eq. \ref{eq:CommMod} and were considered in Ref. \cite{ref:Ali2011}. The best current bound can be obtained from Lamb shift measurements in Hydrogen, which yield $\gamma_0 \lesssim 10^{10}$ (see Table \ref{table:expparamters}).

The strength of the modifications in all the discussed examples depends on the mass of the system. For a harmonic oscillator in its ground state the minimum momentum uncertainty is given by $p_0 = \sqrt{\hbar m \omega_m}$, where $m$ is the mass of the oscillator and $\omega_m$ is its angular frequency. The deformations are therefore enhanced in massive quantum systems. We note that theories of deformed commutators have an intrinsic ambiguity as to which degrees of freedom it should apply to for composite systems (see Supplementary Information). For the center of mass mode, the mass dependence of the deformations suggests that using massive quantum systems allows easier experimental access to the possible deformations of the commutator, provided that precise quantum control can be attained. Opto-mechanical systems, where the oscillator mass can be around the Planck-mass and even larger, therefore offer a natural test-bed for probing commutator deformations of its center of mass mode.

\begin{table}[t]
\centering
\caption{Current experimental bounds on quantum-gravitational commutator deformations. The parameters $\beta_0$ and $\gamma_0$ quantify the deformation strengths of the modification given in Eq. \ref{eq:CommMod} and in Eq. \ref{eq:CommADV}, respectively. For electron tunneling an electric current measurement precision of $\delta I \sim 1\, $fA was taken.}
\begin{tabular}{c c c c}
\\
\hline
system/ experiment         &  $\beta_{0, max}$  &  $\gamma_{0, max}$ & Refs. \vspace{1pt} \\
\hline
\vspace{1pt}
     Position measurement \Tspace &   $10^{34}$  &   $10^{17}$  & \cite{ref:GEO, ref:LIGO} \\
\vspace{1pt}
     Hydrogen Lamb shift    &   $10^{36}$  &   $10^{10}$  & \cite{ref:Das2008, ref:Ali2011}    \\
\vspace{1pt}
     Electron tunneling     &   $10^{33}$  &   $10^{11}$   & \cite{ref:Das2008, ref:Ali2011}
  \\\hline
\end{tabular}
\label{table:expparamters}
\end{table}
%
\section{Scheme to measure the deformations}
\begin{figure}[t]
\includegraphics[width=0.89\columnwidth]{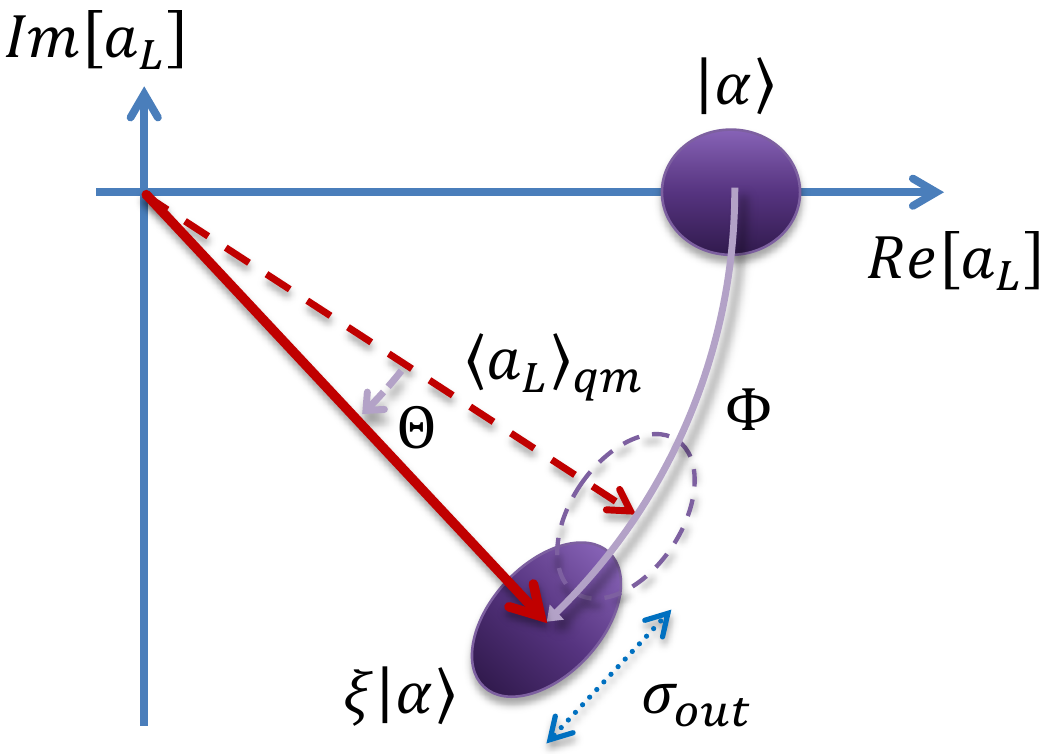}
\caption{\small The effect of the four-displacement operation $\xi$ onto an optical state for the experimentally relevant case $\lambda \ll 1$. In this case, an initial optical coherent state $\ket{\alpha}$ is rotated in phase-space by an angle $\Phi$. A part $\Theta$ of the rotation is due to a possible quantum gravitational deformation of the canonical commutator of the mechanical resonator (see Eq. \ref{eq:MeanLightBeta2}). Measuring the mean of the optical field $\mean{\a}$ and extracting the $\Theta$-contribution allows to probe deformations of the canonical commutator. Optical interferometric schemes can provide a measurement of the overall mean rotation with a fundamental imprecision $\delta \mean{\Phi}=\sigma_{out} /\sqrt{N_p N_r}$, ($N_r$: number of measurement runs, $N_p$: number of photons, $\sigma_{out}$: quadrature width of the optical state, which remains very close to the coherent state value 1/2). In order to resolve the $\Theta$-contribution, the measurement imprecision must fulfill $\delta \mean{\Phi}<\Theta$, which we show can be achieved in quantum opto-mechanical systems even for deformations on the Planck-scale.} \label{fig:Phasespace}
\end{figure}

In the following we will outline a quantum optical scheme that allows to measure deformations of the canonical commutator of a mechanical oscillator with unprecedented precision. For simplicity we use dimensionless quadrature operators $\XM$ and $\PM$. They are related to the position and momentum operators via $x=x_0\XM$ and $p=p_0 \PM$, where $x_0=\sqrt{\hbar/(m \omega_m)}$ and $p_0=\sqrt{\hbar m \omega_m}$.

The scheme relies on displacements of the massive mechanical oscillator in phase space, where the displacement operator is given by \cite{ref:Barnett} $D(z/\sqrt{2})=e^{i(\textrm{Re}[z]\XM - \textrm{Im}[z]\PM)}$. The action of this operator displaces the mean position and momentum of any state by $\textrm{Im}[z]$ and $\textrm{Re}[z]$, respectively. In quantum mechanics, two subsequent displacements provide an additional phase to the state, which can be used to engineer quantum gates \cite{ref:geometric1, ref:geometric2, ref:geometric3}. Here we consider displacements of the mechanical resonator that are induced by an ancillary quantum system, the optical field, with an interaction strength $\lambda$.
A sequence of four opto-mechanical interactions is chosen such that the mechanical state is displaced around a loop in phase space, described by the four-displacement operator
\be
\begin{split}
\xi & = e^{ i \lambda \nL \PM} \,  e^{- i \lambda \nL \XM} \, e^{ -i \lambda \nL \PM} \, e^{ i \lambda \nL \XM} \, .
\end{split}
\ee{eq:FourPulse}
In classical physics, after the whole sequence neither of the two systems would be affected because the four operations cancel each other. However, for non-commuting $\XM$ and $\PM$ there is a change in the optical field depending on the commutator $\comm{\XM}{\PM}=i C_1 $. We can rewrite Eq. \ref{eq:FourPulse} using the well-known relation \cite{ref:Wilcox1967} $e^{a \XM} \PM e^{-a \XM} = \sum_{k=0}^\infty \frac{i^k a^k}{k!} C_k$, where $i C_k = \comm{\XM}{C_{k-1}}$ and $C_0 = \PM$. This yields $\xi = \exp\left(- i \lambda \nL \sum_k \left( \lambda \nL \right)^k C_k/k!  \right)$, which depends explicitly on  the commutation relation for the oscillator, but not on the commutator of the optical field. For the quantum mechanical commutator, i.e. $C_1=1$, we obtain $\xi = e^{-i \lambda^2 \nL^2 }$. In this case, the optical field experiences a self-Kerr-nonlinearity, i.e. an $\nL^2$ operation, and the mechanical state remains unaffected. However, any deformations of the commutator would show in $\xi$, resulting in an observable effect in the optical field.

As an example we consider the modification given by Eq. \ref{eq:CommMod}. To first order in $\beta \equiv \beta_0 \hbar \omega_m m/(M_{P} c)^2 \ll 1$ one obtains $C_1= 1 + \beta \PM^2$, $C_2\approx  \beta  2 \PM$, $C_3 \approx 2 \beta$ and $C_k \approx 0$ for $k \geq 4$. Eq. \ref{eq:FourPulse} thus becomes $\xi_{\beta} = e^{- i \lambda^2 \nL^2 } \, e^{-i \beta \left( \lambda^2 \nL^2 \PM^2  + \lambda^3 \nL^3 \PM + \frac{1}{3} \lambda^4 \nL^4\right)}$ (this approximation has the physical meaning that one can neglect contributions which are higher order in $\beta$ for the observables considered below). As one can see immediately, a deformed commutator affects the optical field differently due to non-vanishing nested commutators $C_k$, $k>1$. In addition to a Kerr-type nonlinearity the optical field experiences highly non-Gaussian $n_L^3$ and $n_L^4$ operations. The additional effect scales with $\beta$ and therefore allows a direct measure of the deformations of the canonical commutator of the mechanical system via the optical field.
To see that explicitly, let us denote the optical field by  $\a$, with the real and imaginary parts representing its measurable amplitude- and phase quadratures, respectively.
Also, for simplicity, we restrict the discussion to coherent states $\ket{\alpha}$ with real amplitudes of the optical input field and we neglect possible deformations in the commutator of the optical field during read-out \cite{ref:Nozari2005, ref:Ghosh2011} since those are expected to be negligible compared to the deformations of the massive mechanical oscillator (see Refs. \cite{ref:parigi2007, ref:Kim2008} for schemes that can probe the non-commutativity of the optical field). For a large average photon number $N_p \gg  1$, and for a mechanical thermal state with mean phonon occupation $\bar{n} \ll \lambda N_p $, the mean of the optical field becomes (for $\abs{\Theta} \ll 1$):
\be
\begin{split}
\mean{\a}  \simeq & \left\langle \a \right\rangle_{qm}  \, e^{- i \Theta}  \, ,
\end{split}
\ee{eq:MeanLightBeta}
where $\mean{\a}_{qm}=\alpha \, e^{ -i \lambda^2 -N_p \left(1- e^{-i 2 \lambda^2} \right)}$ is the quantum mechanical value for the unmodified dynamics. The $\beta$-induced contribution causes an additional displacement in phase space by
\be
\Theta(\beta)  \simeq \frac{4}{3} \beta N_p^3 \lambda^4  \, e^{-i 6 \lambda^2}  \, .
\ee{eq:MeanLightBeta2}
The resulting optical state is represented in Figure 2. We note that the magnitude of the effect is enhanced by the optical intensity and the interaction strength. For the $\mu$- and the $\gamma$-deformation of the commutator, referring to Eqs. \ref{eq:CommMag} and \ref{eq:CommADV}, respectively, the effect on the optical field is similar but shows a different scaling with the system parameters (see Table \ref{table:parameters} and Supplementary Information for the derivation). Probing deviations from the quantum mechanical commutator of the massive oscillator thus boils down to a precision measurement of the mean of the optical field, which can be achieved with very high accuracy via interferometric means, such as homodyne detection.

\section{Experimental implementation}

We now discuss a realistic experimental scenario that can attain sufficient sensitivity to resolve the deformation-induced change in the optical field even for small values of  $\beta_0$, $\mu_0$ and  $\gamma_0$, i.e. in a regime that can be relevant for quantum gravity. The opto-mechanical scheme proposed here can achieve such a regime: it combines the ability to coherently control large masses with strong optical fields. From a more general perspective, opto-mechanical systems provide a promising avenue for preparing and investigating quantum states of massive objects ranging from a few picogram up to several kilogram \cite{ref:review1, ref:review3}. Significant experimental progress has been recently made towards this goal, including laser cooling of nano- and micromechanical devices into their quantum ground state \cite{ref:cooling1, ref:cooling2}, operation in the strong coupling regime \cite{ref:Groblacher2009, ref:OConnell, ref:Teufel2011} and coherent interactions \cite{ref:OConnell, ref:Verhagen2011}. Owing to their high mass they have also been proposed for tests of so-called collapse models \cite{ref:collapse1, ref:revivals, ref:revivals2}, which predict a breakdown of the quantum mechanical superposition principle for macroscopic objects. For our purpose here, which is the high-precision measurement of the canonical commutator of a massive oscillator, we focus on the pulsed regime of quantum opto-mechanics \cite{ref:Vanner2010}.

We consider the setup depicted in Figure 3, where a mechanical oscillator is coupled to the optical input pulse via radiation pressure inside a high-finesse optical cavity. This is described by the intra-cavity Hamiltonian \cite{ref:Law1995} $H = \hbar \omega_m \nM - \hbar g_0 \nL \XM$, where $\nM$ is the mechanical number operator and $g_0 = \omega_c(\xzp/L)$ is the opto-mechanical coupling rate with the mean cavity frequency $\omega_c$ and mean cavity length $L$. For sufficiently short optical pulses the mechanical harmonic evolution can be neglected and the intracavity dynamics can be approximated by the unitary operation $U = e^{i \lambda n_L \XM} $ \cite{ref:Vanner2010}. Here the effective interaction strength is (see Supplementary Information) $\lambda \simeq  g_0/\kappa = 4 \mathcal{F} x_0 / \lambda_L$ , where $\kappa$ is the optical amplitude decay rate, $\mathcal{F}$ is the cavity finesse and $\lambda_L$ is the optical wavelength.
In order to realize the desired displacement operation in phase-space it is also required to achieve a direct opto-mechanical coupling, of the same optical pulse, to the mechanical momentum (see Eq. \ref{eq:FourPulse}). Such a momentum coupling could be achieved for example via the Doppler effect by using mirrors with a strong optical wavelength dependent reflectivity \cite{ref:Karrai2008}. A more straight-forward route is to utilize the harmonic evolution of the mechanical resonator between pulse round-trips (see for example \cite{ref:Vanner2010}), which effectively allows to interchange $\XM$ and $\PM$ after a quarter of the oscillator period.
In this case the contribution from the commutator deformation has a different pre-factor, but remains of the same form (see Supplementary Information), and part of the phase-space rotation in the optical field is of classical nature. This has no effect on the ability to distinguish and to observe the rotation due to the deformed commutator. After the four-pulse interaction has taken place the optical field can be analyzed in an interferometric measurement, which yields the phase information of the light with very high precision.

\begin{figure}[t]
\includegraphics[width=0.96\columnwidth]{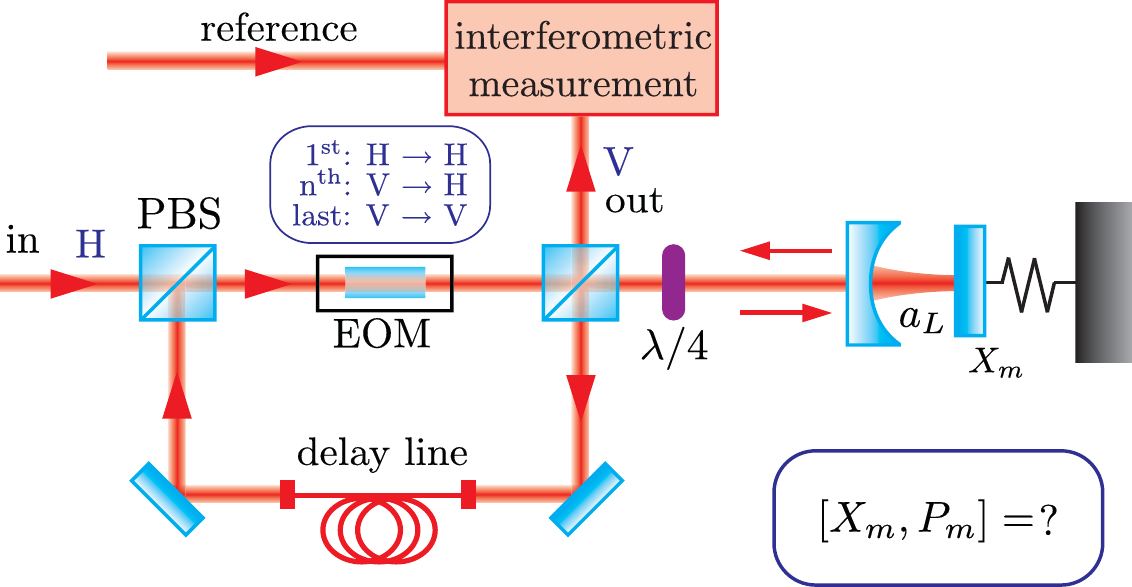}
\caption{\small Proposed experimental setup to probe deformations of the canonical commutator of a macroscopic mechanical resonator. An incident pulse ``in" is transmitted through a polarizing beam splitter (PBS) and an electro-optic modulator (EOM) and then interacts with a mechanical resonator with position $\XM$ via a cavity field $\a$. The optical field is retro-reflected from the opto-mechanical system and then enters a delay line during which time the mechanical resonator evolves for one quarter of a mechanical period. The optical pulse, now vertically polarized, is rotated by the EOM to be horizontally polarized and interacts again with the mechanical resonator. This is repeated for a total of four interactions, such that the canonical commutator of the resonator is mapped onto the optical field. Finally the EOM does not rotate the polarization and the pulse exits in the mode labeled ``out" where it is then measured interferometrically with respect to a reference field and commutator deformations can be determined with very high accuracy.} \label{fig:Setup}
\end{figure}

As in previous approaches to measure possible modifications of the canonical commutator \cite{ref:Das2008, ref:Ali2011}, the relevant question is which ultimate resolution $\delta \beta_0$, $\delta \mu_0$,  $\delta \gamma_0$ the experiments can provide. In case of a null result, these numbers would set an experimental bound for $\beta_0$, $\mu_0$,  $\gamma_0$ and hence provide an important empirical feedback for theories of quantum gravity. We restrict our analysis to the experimentally relevant case $\lambda < 1$, for which the effect of a deformed commutator resembles a pure phase-space rotation of the optical output state by angle $\Phi$ (see Figure 2).
The inaccuracy $\delta \Phi$ of the measurement outcome depends on the quantum-noise $\sigma_{out}$ of the outgoing pulse along the relevant generalized quadrature and can be further reduced by quantum estimation protocols \cite{ref:Boixo2009}. For our purposes we only require to measure the mean optical field, Eq. \ref{eq:MeanLightBeta}. The precision of this measurement is not fundamentally limited and is additionally enhanced by the strength of the field and the number of experimental runs $N_r$ via $\delta\!\mean{\Phi} = \frac{\sigma_{out}}{\sqrt{N_p N_r}}$, from which one directly obtains the fundamental resolutions $\delta \beta_0$, $\delta \mu_0$,  $\delta \gamma_0$. For each of the discussed deformations it is possible to find a realistic parameter regime (see Table \ref{table:parameters}) with dramatically improved performance compared to existing bounds. In particular, we assume a mechanical oscillator of frequency  $\omega_m/2 \pi =10^5$~Hz and mass $m=10^{-11}$~kg, and an optical cavity of finesse $\mathcal{F}=10^5$ at a wavelength of  $\lambda_L=1064$~nm, which is in the range of current experiments \cite{ref:Groblacher2009, ref:Corbitt2007, ref:Harris2008,  ref:Verlot2008, ref:Kleckner2011}. To test a $\mu$-modified commutator (Eq. \ref{eq:CommMag}) a pulse sequence of mean photon-number $N_p=10^8$ is sufficient to obtain a resolution $\delta \mu_0 \sim 1$ already in a single measurement run ($N_r=1$). For the case of a $\gamma$-modified commutator (Eq. \ref{eq:CommADV}) the same sequence would result in $\delta \gamma_0 \sim 10^9$. By increasing the photon-number to $N_p= 5 \times  10^{10}$, the Finesse to $\mathcal{F}=2 \times 10^5$  and the number of measurement runs to $N_r=10^5$ (this would require to stabilize the experiment on a timescale of the order of seconds) one obtains $\delta \gamma_0 \sim 1$. Note that this would improve the existing bounds for $\gamma_0$ \cite{ref:Ali2011} by 10 orders of magnitude. To obtain similar bounds for a $\beta$-modification is more challenging. The pulse sequence with the previous parameters yields $\delta \beta_0 \sim 10^{12}$, which already constitutes an improvement by about 20 orders of magnitude compared to the current bound for $\beta_0$ \cite{ref:Das2008}. This can provide experimental access to a possible intermediate length-scale or a meaningful feedback to theories of quantum gravity in the case of a null result. By further pushing the parameters to $N_p= 10^{14}$, $N_r=10^6$, $\mathcal{F}= 4 \times 10^5$, $m =10^{-7}$~kg and $\lambda_L=532$~nm it is even possible to reach $\delta \beta_0 \sim 1$, i.e. a regime where Planck-scale deformations are relevant and 33 orders of magnitude beyond current experiments. To achieve such experimental parameters is challenging, but is well within the reach of current technology.

The above considerations refer to the ideal case in which experimental noise sources can be neglected. Effects such as mechanical damping and distortions of the effective interaction strength impose additional requirements on the experimental parameters, which are discussed in detail in the Supplementary Information.
In summary, being able to neglect the effects of pulse shape distortion and optical loss requires that the mechanical mode is optically cooled close to about a thermal occupation of $\bar{n} < 30$. Similarly, decoherence effects are negligible when the whole mechanical system is in a bath of temperature $T < 100$~mK for resonators with a quality factor of $Q > 10^6$, which can be achieved with dilution refrigeration. In general, the scheme is very robust against many noise sources, since it relies on the measurement of the mean of the optical field and the noise sources can be isolated by independent measurements. We also note that contributions from a modified commutator scale in a different way with the system parameters as compared to deleterious effects. It is therefore possible to distinguish these by varying the relevant parameters, such as optical intensity and the oscillator's mass.
The proposed scheme thus offers a feasible route to probe possible effects of quantum gravity in a tabletop quantum optics experiment and hence to provide important empirical feedback for theories of quantum gravity.

\begin{table}[t]
\renewcommand{\arraystretch}{1.15}
\centering
\caption{Experimental parameters to measure quantum gravitational deformations of the canonical commutator. The parameters are chosen such that a precision of $\delta \mu_0 \sim 1$, $\delta \gamma_0 \sim 1$ and $\delta \beta_0 \sim 1$ can be achieved, which amounts to measuring Planck-scale deformations.}
\begin{tabular}{c |c |c |c }
$\comm{x}{p}$& Eq. \ref{eq:CommMag}  & Eq. \ref{eq:CommADV} &  Eq. \ref{eq:CommMod} \\
$|\Theta|$  & $ \frac{\mu_0 32 \hbar \mathcal{F}^2  m N_p}{M_{P}^2 \lambda_L^2 \omega_m} $ & $ \frac{\gamma_0 96 \hbar^2 \mathcal{F}^3  N_p^2}{ M_{P}c \lambda_L^3 m \omega_m} $ & $ \frac{\beta_0 1024 \hbar^3 \mathcal{F}^4 N_p^3}{3 M_{P}^2 c^2 \lambda_L^4 m \omega_m } $ \TspaceBig \Bspace \\
\hline
$\mathcal{F}$    &   $10^{5}$            & $2 \times 10^{5}$    &  $4 \times 10^{5}$ \Tspace  \\
$m$              &   $ 10^{-11}$~kg      & $ 10^{-9}$~kg       & $ 10^{-7}$~kg  \\
$\frac{\omega_m}{2\pi}$&   $10^{5}$~Hz         & $10^{5}$~Hz          & $ 10^{5}$~Hz  \\
$\lambda_L$      &   $1064$~nm           & $1064$~nm            & $532$~nm  \\
$N_p$            &   $10^{8}$            & $5 \times 10^{10}$   & $ 10^{14}$ \\
$N_r$            &   $1$                 & $10^5$               & $10^6$ \\
$\delta\!\mean{\Phi}$& $10^{-4}$         & $ 10^{-8}$   & $10^{-10}$ \\
\end{tabular}
\label{table:parameters}
\end{table}
%

\noindent
\\
\textbf{Acknowledgements}
\noindent \\
This work was supported by the Royal Society, by the EPSRC, by the European Comission (Q-ESSENCE), by the European Research Council (ERC QOM), by the Austrian Science Fund (COQUS, START, SFB FOQUS), by the Foundational Questions Institute and by the John Templeton Foundation. M.R.V. is a recipient of a DOC fellowship of the Austrian Academy of Sciences. I.P. and M.R.V are members of the FWF Doctoral Programme CoQuS and they thank the kind hospitality provided by Imperial College London. The authors thank S. Das, S. Gielen, H. Grosse, A. Kempf, M. Plenio, A. Retzker, W. T. Kim and J. Lee for discussions.

\pagebreak
\section*{Supplementary Information}

\begin{appendix}

\section{Alternative Commutator deformations}
In this section we compute the change in the optical field for the two alternative commutator deformation theories considered in the main text, i.e. for the $\mu$- and the $\gamma$-deformation given by Eqs. \ref{eq:CommMag} and \ref{eq:CommADV}, respectively. For the case $\comm{x}{p}_{\mu_0} \approx i \hbar \left( 1+ \mu_0 m^2 / M_{P}^2\right)$ the four-displacement operator becomes $\xi_{\mu} = e^{- i \lambda^2 \nL^2 \left( 1 + \mu \right)}$, where we defined $\mu \equiv \mu_0 m^2/M_{P}^2$. The resulting mean of the optical field is thus $\mean{\a}  =  \alpha  e^{ -i \lambda^2 \left( 1 + \mu \right)} \, e^{-N_P \left(1- e^{-i 2 \lambda^2 \left( 1 + \mu \right)}\right) }$. In the limit $\mu \lesssim 1$ and $N_P \gg 1$, it reduces to $\mean{\a}  \simeq  \left\langle \a \right\rangle_{qm}  \, e^{- i \Theta}$ with the deformation-induced contribution $\Theta(\mu)$ given by
\begin{equation*}\label{eq:MeanLightMag2}\tag{A.1}
\Theta(\mu) \simeq 2 \mu N_p \lambda^2  e^{-i 2 \lambda^2} \, .
\end{equation*}

For a $\gamma$-deformation of the commutator we define $\gamma \equiv \gamma_0 \sqrt{\hbar m \omega_m} / M_{P}c \ll 1$ and the four-displacement operator becomes  $\xi_{\gamma} = e^{- i \lambda^2 \nL^2 } \, e^{i \gamma \left( \lambda^2 \nL^2 \PM  + \frac{1}{2} \lambda^3 \nL^3 \right)}$ to first order in $\gamma$. In the limit $N_P \gg 1$, this results in the additional contribution to the quantum mechanical mean of the optical field given by
\begin{equation*}\label{eq:MeanADV2}\tag{A.2}
\Theta(\gamma)  \simeq \frac{3}{2} \gamma N_p^2 \lambda^3  \, e^{-i 4 \lambda^2}  \, .
\end{equation*}

Finally, we note that theories with a modified commutator have an intrinsic ambiguity as to which particles or degrees of freedom of a composite system the deformations should apply to. For example, a system consisting of $N$ particles with identical mass and each with position $x_i$ and momentum $p_i$ has the center of mass degrees of freedom given by $x_{cm}=\sum_i^{N} x_i/N$ and $p_{cm}= \sum_i^{N} p_i$. If the $\beta$-modified commutation relation as given by Eq. \ref{eq:CommMod} is applied to each single particle individually (rather than to the center of mass itself), the commutator of the center of mass becomes $\comm{x_{cm}}{p_{cm}}= i \hbar \left( 1+ \beta_0 \sum_i^{N} p_i^2/( M_{P}^2 c^2 N) \right) =  i \hbar \left( 1+ \frac{\beta_0}{M_{P}^2 c^2 N} \left( p_{cm}^2 -  \sum_{i \neq j}^{N} p_i p_j \right) \right)$. This result differs from a direct deformation of the center of mass mode. Depending on the state of the system, in particular on the pairwise correlations between the constituent particles, the difference in the commutator deformation can be approximated by the substitution $ \beta_0 \rightarrow \chi \beta_0 $ where $\chi$ lies between $1/N$ (for vanishing pairwise correlations) and $1/N^2$ (for all pairs being equally correlated).

\section{Modified harmonic evolution}

For the implementation utilizing free harmonic evolution between the pulsed interactions, it is necessary to take a deformed evolution due to a deformation of the commutator into account. Here we consider the $\beta$-deformation as given by Eq. \ref{eq:CommMod}. Keeping the quantum mechanical canonical commutation relation and modifying the momentum operator to $\PM \rightarrow \PM \left(1 + \frac{1}{3} \beta \PM^2 \right)$ incorporates the deformation to first order in $\beta$ \cite{ref:Das2008}. It is therefore necessary to solve for $\XM(t)$ for a modified evolution governed by the effective Hamiltonian $H= \frac{1}{2}\hbar \omega_m \left( \XM^2 + \PM^2 \right) + \frac{1}{3}\hbar\omega_m \beta \PM^4 \equiv H_0 + H'$. These correspond to the unitary evolutions $U_0$ and $U'$, respectively. In a frame rotating at frequency $\omega_m$, the time evolution of the operators $\XM'= U_0 \XM U_0^{\dagger}$ and $\PM'=U_0 \PM U_0^{\dagger}$ is generated by $H'(\PM')$. This yields $\PM'(t)=\PM'(0)$ and $\XM'(t)=\XM'(0) + \frac{4}{3} \beta \omega_m t \PM'^3$. In the original frame, the result is thus
\begin{equation*} \tag{B.1}
\begin{split}
& \XM(t)  = \XM(0) \cos(\omega_m t) - \PM(0) \sin(\omega_m t) \, + \\
& + \frac{4}{3} \beta \omega_m t \left( \PM(0) \cos(\omega_m t) + \XM(0) \sin(\omega_m t) \right)^3 \, .
\end{split}
\end{equation*}
Using four interactions separated by a quarter mechanical period, the four-displacement operator becomes $\xi=e^{i \lambda \nL (\PM - 2 \beta \pi \XM^3)}e^{i \lambda \nL (-\XM - 4 \beta \pi \PM^3/3)} \times e^{i \lambda \nL (-\PM + 2 \beta \pi \XM^3/3)} e^{i \lambda \nL \XM}$. This expression can be simplified using the Zassenhaus formula \cite{ref:Wilcox1967} $\exp(X+Y)=\exp(X)\exp(Y) \prod_{i=1}^{\infty}\exp(Z_i)$, where $Z_1=-\comm{A}{B}/2$, $Z_2=\comm{A}{\comm{A}{B}}/6 + \comm{B}{\comm{A}{B}}/3$, $Z_3=-\left( \comm{B}{\comm{A}{\comm{A}{B}}} + \comm{B}{\comm{B}{\comm{A}{B}}}\right)/8 -\comm{A}{\comm{A}{\comm{A}{B}}}/24$ and $Z_k$, $k>3$ are functions of higher nested commutators.
To leading order in $\nL$, the four-displacement operator becomes
\begin{equation*} \tag{B.3}
\xi \simeq e^{-i \lambda^2 \nL^2} \, e^{i \beta \pi \frac{5}{3} \lambda^4 \nL^4} .
\end{equation*}
The optical field due to this operation is of the same form as in Eq. \ref{eq:MeanLightBeta} with a modified numerical strength. The modified dynamics therefore does not alter the main conclusions.

\section{Additional requirements due to deleterious effects} \label{App3}

In the following we analyze the experimental parameters necessary to overcome some additional deleterious effects in the opto-mechanical system. We analyze the cavity dynamics and its influence on the effective interaction, the effect of varying interaction strengths for each pulse round trip and the influence of mechanical decoherence.  We neglect additional contributions from a modified commutator since these will be less prominent than that considered in the ideal case.

The Hamiltonian $H = \hbar \omega_m \nM - \hbar g_0 \nL \XM$ refers to the interaction between the optical field and the mechanics within the cavity \cite{ref:Law1995}. To quantify the effects of cavity filling and decay for a short pulse we solve the optical Langevin equation
\be
\frac{d\a}{dt} = (i g_0 \XM - \kappa) \a + \sqrt{2 \kappa} \left( \a^{(in)} + \sqrt{N_p} \alphai \right)
\ee{eq:cavity}
with the boundary condition $\a^{(in)}+ \a^{(out)}= \sqrt{2 \kappa} \a$ for the input and output optical fields and the incident cavity drive $\alphai$ that is normalized to the mean photon number per pulse, i.e. $\integ{t} \alphai^2 = 1$. Since the mechanical motion can be neglected in the short pulse regime the overall effect on both the optical field and the mechanical oscillator can be described by the effective unitary operator $U = e^{i \lambda \nL \XM}$. The coupling strength $\lambda$ depends on the intra-cavity field envelope and can be determined via the total momentum transfer onto the mechanics by the optical pulse $\mean{\PM}~=~g_0 \int dt \mean{\nL(t)}$, where $\nL(t)$ is obtained from Eq. \ref{eq:cavity}. This yields  $\lambda = \zeta \, g_0 / \kappa$ with $\zeta = \int \! \textrm{d}t e^{-2\kappa t} \kappa^2 \left[ \integlim{-\infty}{t}{t'} e^{\kappa t'} \alphai(t') \right]^2$ for the effective unitary operator.

In general, the pulse shape of the output optical field is altered by the cavity.
When such a distorted pulse is directed back for the $i$-th time into the cavity, the effective interaction time within the cavity will be different and will give rise to a modified opto-mechanical interaction strength $\lambda_i$. To minimize the distortion, one requires that the pulse duration $\tau$ is much longer than the intra-cavity lifetime, i.e. $\omega_m \ll \tau^{-1} \ll \kappa$, where $\kappa$ is the cavity bandwidth. This ensures that the optical pulses are short compared to the mechanical period and that the cavity is empty in between the pulsed interactions. In this regime we have $\zeta \simeq 1$ such that $\lambda \simeq g_0 / \kappa$.

An additional effect that distorts the interaction strength $\lambda$ from pulse to pulse is the loss of light.
To include both loss and pulse shape change in the effective interaction, we define an overall distortion parameter $\eta$. With this parameter, the opto-mechanical interaction strength $\lambda_i$ for the $i$-th interaction is approximately given by $\lambda_{i+1} = \eta \lambda_i $. We note that in the regime considered here the loss of light will be dominant and we assume a value of $\eta \sim 0.9$.

The effect of varying interaction strengths modifies the four-displacement operator to $\xi_{\eta} = e^{ i \lambda_4 \nL \PM} \times e^{- i \lambda_3 \nL \XM}  e^{ -i \lambda_2 \nL \PM} e^{ i \lambda_1 \nL \XM}  $. Using $\lambda=\lambda_1$, it can be written as
\be
\xi_{\eta}  = \xi'_0 \, e^{i \eta \lambda(1-\eta^2) \nL \PM} \, e^{i \lambda(1-\eta^2) \nL \XM} \, ,
\ee{eq:FourPulseEpsilon}
where $\xi'_0$ is the four-displacement operator as considered previously,  but with modified interaction strengths: For the $\beta$-, $\gamma$- and $\mu$-deformations, the interaction strength is reduced to $\lambda^4 \rightarrow \eta^7 \lambda$, $\lambda^3 \rightarrow \eta^5 \lambda$ and $\lambda^2 \rightarrow \eta^3 \lambda$, respectively. For $\eta \sim 0.9$ the $\Theta-$contribution to the optical mean by the $\beta$-modified commutator would therefore be reduced by a factor $\sim 0.5$, the contribution by the $\gamma$-modified commutator would be reduced by $\sim 0.6$ and the contribution by a $\mu$-modified commutator would be reduced by $\sim 0.7$. Additionally, Eq. \ref{eq:FourPulseEpsilon} contains a strong dependence of the outgoing optical field on the mechanical state. Given a thermal distribution of the mechanical center-of-mass mode with mean phonon occupation $\nbar$, the optical mean is reduced by $e^{-\nbar \lambda^2(1-\eta^2)(1-\eta^4)/2}$. For $\eta \sim 0.9$ and $\lambda \sim 1$, the mechanics therefore needs to be damped to $\nbar \lesssim 30 $. This can be achieved by optical cooling of the mechanical mode, which has recently been demonstrated in Refs. \cite{ref:cooling1, ref:cooling2}.

Finally, we discuss mechanical decoherence in between pulse interactions due to coupling of the mechanical mode to other degrees of freedom in the oscillator. We consider a linear coupling to an infinite bath of harmonic oscillators, which can be described by the interaction Hamiltonian
\be
H_{int} = \sum_i \nu_i  \left( b_i \, e^{-i \omega_i t} + b_i^{\dagger} \, e^{i \omega_i t} \right) \XM \, ,
\ee{}
where $b_i$ are operators for the $i$-th bath mode with frequency $\omega_i$ that interact with the mode of interest with interaction strength $\nu_i$. Using the notation $B(t) = \sum_i \nu_i  \left( b_i \, e^{-i \omega_i t} + b_i^{\dagger} \, e^{i \omega_i t} \right)$, the solutions for the position and momentum  operators become
\be
\begin{split}
\XM(t) & = \XM^{(0)}(t,t_0) - \integlim{t_0}{t}{t'} \! \! B(t') \sin(\omega_m (t-t'))
\\
\PM(t) & = \PM^{(0)}(t,t_0) + \integlim{t_0}{t}{t'} \! \! B(t') \cos(\omega_m (t-t')) \, ,
\end{split}
\ee{}
where $\XM^{(0)}(t,t_0)= \textrm{Re}[A(t_0) e^{i \omega_m (t-t_0)}]$ and $\PM^{(0)}(t,t_0)= \textrm{Im}[A(t_0) e^{i \omega_m (t-t_0)}]$ are the position and momentum operators without decoherence, respectively, with the initial value $A(t_0) = \XM(t_0) + i \PM(t_0)$. For a bath that is initially uncorrelated with the mechanical mode of interest, the $\xi$-operator changes to
\be
\xi_B =  \xi_0 \, e^{i \lambda \nL B_3} \, e^{i \lambda \nL B_2} \, e^{i \lambda \nL B_1}
\ee{}
where $\xi_0$ is the operator without decoherence as given in Eq. \ref{eq:FourPulse} and the bath degrees of freedom enter through the operators $B_1~=~\int_0^{\pi/2 \omega_m} dt \! B(t') \cos(\omega_m t)$, $B_2~= \int_0^{\pi/ \omega_m} dt \! B(t') \sin(\omega_m t)$ and $B_3~= -~\int_0^{3 \pi/2 \omega_m} dt \! B(t') \cos(\omega_m t)$. We consider a Markovian bath with negligible bath correlation times such that $\mean{B(t)}=0$ and $\mean{B(t) B(t')}=\gamma_m \coth(\hbar \omega_m /2 k_B T) \delta(t-t')$, where the mechanical damping can be written in terms of the mechanical quality factor as $\gamma_m= \omega_m / Q$. To first order in $T/Q$ the mean of the optical field becomes
\be
\mean{\a}_{B} \simeq \mean{\a}_0 (1 - \lambda^2 \frac{k_B T}{\hbar \omega_m Q}) \, ,
\ee{}
where $\mean{\a}_0$ is the mean of the optical field without decoherence. For $Q=10^6$ one therefore requires $T \lesssim 100$~mK to keep the decoherence sufficiently weak. Such parameters can be achieved for kHz-resonators with dilution refrigeration.

\end{appendix}

\end{document}